\documentclass[11pt,twocolumn]{article} 

\usepackage{times}
\usepackage{epsf}
\usepackage{psfig}
\newcommand{\ls}[1]
   {\dimen0=\fontdimen6\the\font
    \lineskip=#1\dimen0
    \advance\lineskip.5\fontdimen5\the\font
    \advance\lineskip-\dimen0
    \lineskiplimit=.9\lineskip
    \baselineskip=\lineskip
    \advance\baselineskip\dimen0
    \normallineskip\lineskip
    \normallineskiplimit\lineskiplimit
    \normalbaselineskip\baselineskip
    \ignorespaces
   }

\newcounter{LineNum}
\newcounter{AlgNum}

\newsavebox{\savepar}

\newcommand{\pscript}[2]
{\setlength{\epsfxsize}{#2\hsize}
 \centerline{\epsfbox{#1}}}

\long\def\omitit#1{}
\begin{document}
\ls{1.0}

\title{On the Tomography of Networks and Multicast Trees}


\author{
  Reuven Cohen 
  \\ inerva Center and Department of Physics \\ Bar-Ilan university \\ Ramat-Gan, Israel \\ {\tt cohenr@shoshi.ph.biu.ac.il}
  \and
  Danny Dolev 
  \\  School of Engineering and Computer Science\\ Hebrew University \\ Jerusalem, Israel \\ {\tt dolev@cs.huji.ac.il} \\
  \and 
  Shlomo Havlin
  \\ inerva Center and Department of Physics \\ Bar-Ilan university \\ Ramat-Gan, Israel \\ {\tt havlin@ophir.ph.biu.ac.il}
  \and
  Tomer Kalisky
  \\ inerva Center and Department of Physics \\ Bar-Ilan university \\ Ramat-Gan, Israel \\ {\tt kaliskt@mail.biu.ac.il}
  \and
  Osnat Mokryn 
  \\  School of Engineering and Computer Science \\ Hebrew University \\ Jerusalem, Israel \\ {\tt osnaty@cs.huji.ac.il}
  \and
  Yuval Shavitt
  \\  Dept. of Electrical Engineering - Systems \\ Tel-Aviv University \\  Tel-Aviv, Israel \\ {\tt shavitt@ieee.org}
}

\maketitle

\begin{abstract}
In this paper we model the tomography of scale free networks by
studying the structure of layers around an arbitrary network node. We
find, both analytically and empirically, that the distance
distribution of all nodes from a specific network node consists of two
regimes. The first is characterized by rapid growth, and the second
decays exponentially.  We also show that the nodes degree distribution
at each layer is a power law with an exponential cut-off.  We obtain
similar results for the layers surrounding the root of multicast trees
cut from such networks, as well as the Internet.  All of our results
were obtained both analytically and on empirical Interenet data.
\end{abstract}
\section{Introduction} 
\label{sec:intro}

In recent years there is an extensive effort to model the topology of
the Internet.  While the exact nature of the Internet topology is in
debate~\cite{CC+02}, it was found that many realistic networks posses
a power law, or scale free degree distribution~\cite{FFF99}. These
results were also verified by~\cite{GT00,MMB00,CNSSV99,DMS03}, who
conducted further investigations.  Albert and
Barab\'{a}si~\cite{BA99,AB00} suggested a dynamic graph generation
model that generates such networks. One of their main findings was the
self similarity characteristic of such networks.  Interestingly,
empirical findings on partial views obtained similar results, which
may lead to the assumption that due to the self similarity nature of
the Internet structure, this characteristic would be exposed through
different cuts and filters.

In this paper we study the tomography of scale free networks and
multicast trees embedded on them.
We use the Molloy Reed graph generation method~\cite{MR98} in conjunction
with similar techniques to study the layer structure (tomography) of
networks.  Specifically, we study the number and degree distribution
of nodes at a given (shortest path) distance from a chosen network node.  We
show analytically that the distance distribution of all nodes from a
specific network node consists of two regimes. The first can be
described as a very rapid growth, while the second is found to decay
exponentially.  We also show that the node degree distribution at each
layer obeys a power law with an exponential cut-off.  We back our
analytical derivations with simulations, and show that they match.

We also study shortest path trees cut from scale free networks, as they
may represent multicast trees. We investigate their layer structure
and distribution.  We show that the structure of a
multicast tree cut from a scale free network exhibits a layer behavior
similar to the network it was cut from. We validate our analysis
with simulations and real Internet data.

The paper is organized as follows. Section~\ref{sec:related} details
previous findings and gives the basic terminology we use in the
paper. In Section~\ref{sec:net_anal} we introduce the process used for
generating scale free graphs and their layers. Then, we analyze the
resulting tomography of such networks, and back the results with
simulations and real data.  In Section~\ref{sec:trees} we investigate
the tomography of multicast trees cut from such networks, and back our
findings with real Internet data.

\section{Background}
\label{sec:related}
\subsection{Graph Generation}
In recent years studies have shown that many real world networks, and, in
particular, the Internet, are scale free networks. That is, their degree
distribution follows a power law, $P(k)=ck^{-\lambda}$, where $c$ is an
appropriate normalization factor, and $\lambda$ is the exponent of the power
law.

Several techniques for generating such scale free graphs were
introduced~\cite{BA99,MR98}.  Molloy and Reed suggested in~\cite{MR98}
a method (MR model) for computing the size of the
\omitit{biggest}giant (or largest) component in a scale free
network. To do so, they developed the following method.  A graph with
a given degree distribution is generated out of the probability space
(ensemble) of possible graph instances.  For a given graph size $N$,
the degree sequence is determined by randomly choosing a degree for
each of the $N$ nodes from the degree distribution.\omitit{Thus each node is
given a set of ``open'' outgoing links which are not connected yet to
another node.}  Let us define $V$ as the set of $N$ chosen nodes, $C$
as the set of unconnected outgoing links from the nodes in $V$, 
and $E$ as the set
of edges in the graph. Initially, $E$ is empty.  Then, the links in
$C$ are randomly matched, such that at the end of the process, $C$ is
empty, and $E$ contains all the matched links $<u,v>$, $u,v \in V$.
Throughout this paper, we refer to the set of links in $C$ as open
connections.

Note, that while in the BA model the graph degree distribution
function exists only at the end of the process, in the MR model the
distribution is known apriori, thus enabling us to use it in our
analysis during the construction of the graph.

\subsection{Cut-Off Effect}
Recent work~\cite{CH02}, has shown that the radius
\footnote{ We define the radius of a graph, $r$, as the average
distance of all nodes in the graph from the node with the highest
degree (if there is more than one we will arbitrarily
choose one of them). The average hop distance or diameter of the
graph, $d$, is restricted to:
\begin{equation}
r\leq d\leq 2r,
\end{equation}
Thus the average hop sequence is bound from above and from below by
the radius. }
, $r$, of scale free graphs is extremely small and scales as $r \sim \log
\log N$. The meaning of this is that even for very large networks,
finite size effects must be taken into account, because algorithms
for traversing the graph will get to the network edge after a small number of steps.

Since the scale free distribution has no typical scale, its behavior is
influenced by externally imposed cutoffs, i.e. minimum and maximum values for 
the allowed degrees, $k$. The fraction of sites having degrees above and below
the threshold is assumed to be $0$. The lower cutoff, $m$, is usually chosen to
be of order $O(1)$, since it is natural to assume that in real world networks
many nodes of interest have only one or two links. The upper cutoff, $K$, can
also be enforced externally (say, by the maximum number of links that can be
physically connected to a router). However, in situations where no such cutoff
is imposed, we assume that the system has a natural cutoff.

To estimate the natural cutoff of a network, we assume that the network consists
of $N$ nodes, each of which has a degree randomly selected from the distribution
$P(k)=ck^{-\lambda}$. An estimate of the average value of the largest of the $N$
nodes can be obtained by looking for the smallest possible tail that contains 
a single node on the average~\cite{CEBH00}:
\begin{equation}
\sum_{k=K}^\infty P(k)\approx \int_K^\infty P(k)dk=1/N.
\end{equation}
Solving the integral yields $K\approx mN^{1/(\lambda-1)}$, which is the
approximate natural upper cutoff of a scale free network~\cite{CEBH00,DM01,MAA02}. 

In the rest of this paper, in order to simplify the analysis
presented, we will assume that this natural cutoff is imposed on the
distribution by the exponential factor $P(k)=ck^{-\lambda}e^{-k/K}$. 

\section{Tomography of Scale Free Networks}
\label{sec:net_anal}

In this section we study the statistical behavior of chemical layers
surrounding the maximally connected node in the network.  
First, we describe the process of generating the network, and define 
our terminology. Then, we analyze the degree distribution at each layer
surrounding the maximally connected node.
\omitit{
In particular we will use the Molloy-Reed ~\cite{MR98} model for scale
free networks, and construct or ``expose'' the network following the
method introduced in ~\cite{CH02}.
}

\subsection{Model Description}
We base our construction on the Molloy-Reed model~\cite{MR98}, 
also described in~\ref{sec:related}.
The construction process tries to gradually expose the network, following 
the method introduced in~\cite{CH02}, and is forcing a hierarchy on the
Molloy-Reed model, thus enabling us to define layers in the graph.

We start by setting the number of nodes in the network, N. We then choose 
the nodes degrees according to the scale-free distribution function
$P(k)=ck^{-\lambda}$, where $c \approx (\lambda-1)m^{\lambda-1}$ is
the normalizing constant and $k$ is in the range $[m,K]$, for some
chosen minimal degree $m$ and the natural cutoff $K=mN^{1/(\lambda-1)}$ of the 
distribution~\cite{CEBH00,DM01}.

At this stage each node in the network has a given number of outgoing
links, which we term {\em open connections}, according to its chosen
degree.  Note, that according to the terminology in~\ref{sec:related},
the set of links in $E$ is empty at this point, while the set of
outgoing open links in $C$ contains all unconnected outgoing links in
the graph.  We proceed as follows: we start from the maximal degree
node, which has a degree $K$, and connect it randomly to $K$ available
open connections, thus removing these open connections from $C$.  We
have now exposed the first {\em layer} (or {\em shell}), indexed as
$l=1$. We now continue to fill out the second layer $l=2$ in the same
way: We connect all open connections emerging from nodes in layer
No. $1$ to randomly chosen open connections. These open connections
may be chosen from layer No. $1$ (thus creating a loop) or from othe
links in $C$. We continue until all open connections emerging from
layer No. $1$ have been connected, thus filling layer $l=2$.
Generally, to form layer $l+1$ from an arbitrary layer $l$, we
randomly connect all open connections emerging from $l$ to either
other open connections emerging from $l$ or chosen from the othe links
in $C$. Note, that when we have formed layer $l+1$, layer $l$ has no
more open connections.  The process continues until the set of open
connections, $C$, is empty.

\omitit{
\subsection{Symbols}
We define the following symbols:

\begin{itemize}
\item $N$ - The number of nodes in the network.
\item $P(k)=ck^{-\lambda}$ - The degree distribution of nodes in the
network, where $c \approx (\lambda-1)m^{\lambda-1}$ is the normalizing
constant, k is in the range $[m,K]$, m is some chosen minimal degree
and $K$ is the maximal degree. If no upper cutoff is imposed, we take
$K=mN^{1/(\lambda-1)}$.
\item $S_l$ - The number of incoming connections into layer $l$. This
may be considered a good approximation for the number of nodes in
layer No. $l$ in cases when each site in layer $l$ is reached by a
single incoming connection.
\item $\chi_l$ - The number of open connections emerging from layer $l$. 
\item $P_l(k)$ - The degree distribution of all nodes remaining
outside layer No. $l$, or the probability of a node to be outside
layer No. l and to have k connections.
\item $T_l$ - The total number of open connections outside layer
No. $l$ (that is, remaining not connected after filling layer No. $l$
and before exposing its outgoing connections).
\item $K_l$ - The upper cutoff of $P_l(k)$, the distribution outside
layer No. $l$. We will prove that at each layer the distribution
$P_l(k)$ is equivalent to the original network distribution,
multiplied by an exponential cutoff $e^{-k/K_l}$.
\end{itemize}
}

\begin{figure}
\epsfxsize=2.8in
\hskip 0.05in
\epsfbox{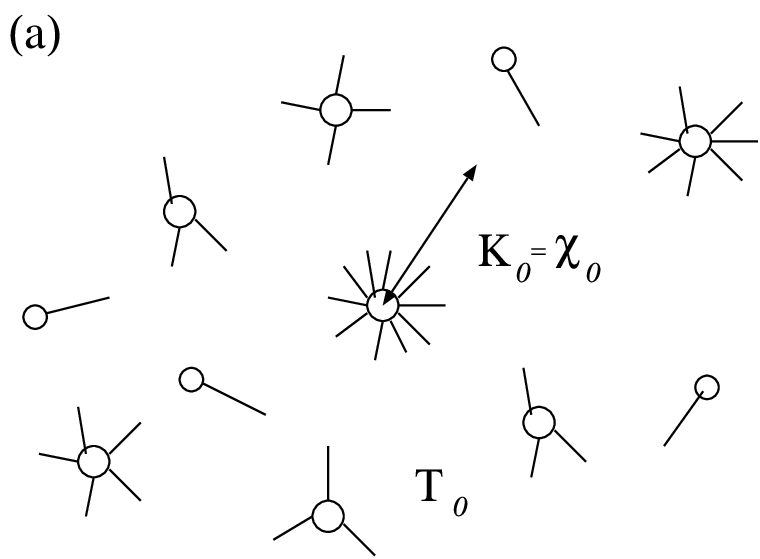}
\epsfxsize=3in
\hskip 0.05in
\epsfbox{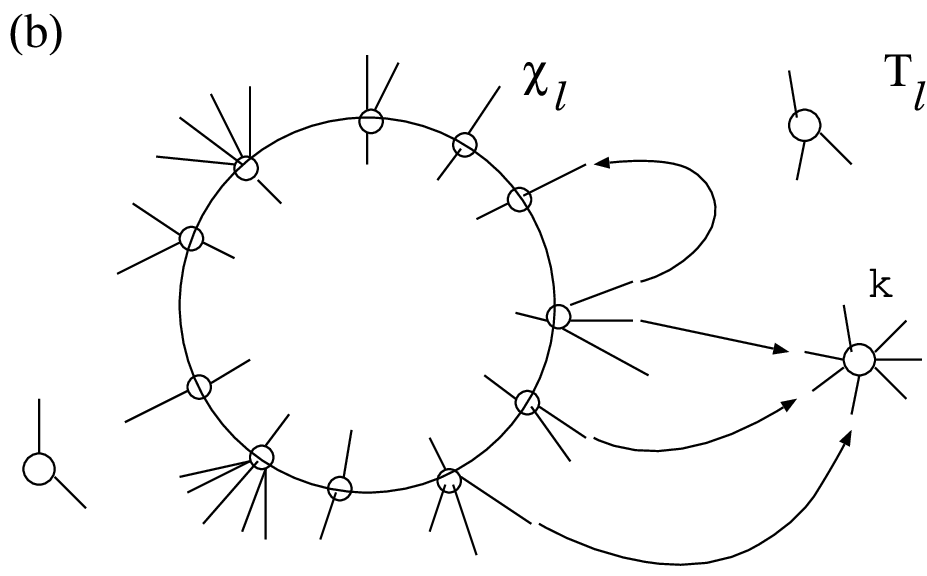}
\epsfxsize=3in
\hskip 0.05in
\epsfbox{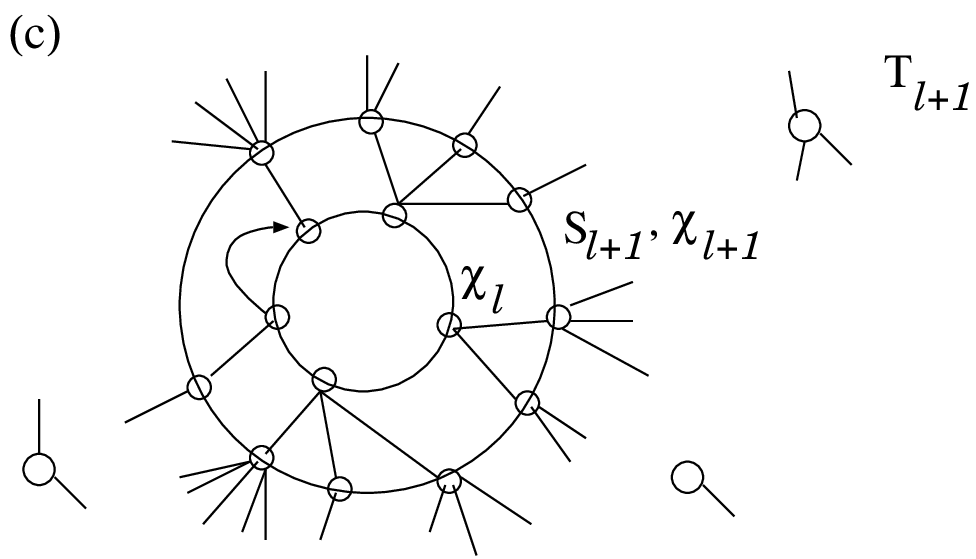}
\vskip 0.2in
\caption{\small Illustration of the exposure process.  The large
circles denote exposed layers of the giant component, while the small
circles denote individual sites.  The sites outside the circles have
not been reached yet.  (a) We begin with the highest degree node and
fill out layer No.$1$.  (b) In the exposure of layer No. $l+1$ any
open connection emerging from layer No. $l$ may connect to any open
node ($T_l$ connections) or loop back into layer No. $l$ ($\chi_l$
connections).  (c) The number of connections emerging from layer
No. $l+1$ is the difference between $T_l$ and $T_{l+1}$ after reducing
the incoming connections $S_{l+1}$ from layer No. $l$. }
\label{illust} 
\end{figure}

\subsection{Analysis}
\omitit{
We begin with the following values:
\begin{itemize}
\item $K_0=K$ - we begin with the natural cutoff of the graph.
\item $\chi_0=K_0$ - number of open connections emerging from the
maximal degree node (layer No. $0$).
\item $P_0(k)=ck^{-\lambda}e^{-k/K_0}$ - the degree distribution of
all nodes remaining outside the maximal degree node (layer
No. $0$). Note that we changed the natural network cutoff $K_0$ with
an exponential cutoff. This will be justified for consecutive layers.
\end{itemize}
}

We proceed now to evaluate the probability for nodes with degree $k$ to reside
in any of the layers layer $\{i|i>l\}$ for some $l$, denoted by $P_l(k)$.

The number of open connections outside layer No. $l$, is given by:

\begin{equation}
\label {T_l}
T_l=N\sum_k kP_l(k)
\end {equation}

Thus, we can define the probability that a detached node with degree $k$ will
be connected to an open connection emerging from 
layer $l$ by \(\frac{k}{\chi_l+T_l}\),
where $\chi_l$ is the number of open connections emerging from layer $l$.

Therefore, the conditional probability for a node with degree $k$ 
not to be also in layer $l+1$, given that it cannot connect to {\em any} of
the $\chi_l$ open connection emerging from layer $l$, is:
\begin{eqnarray}
\label {conditional_Prob_to_be_outside_l+1}
P(k,l+1|l)=\left [1-\frac{k}{\chi_l+T_l} \right ]^{\chi_l} \approx
\nonumber \\ \approx exp \left(-\frac{k}{1+\frac{T_l}{\chi_l}} \right)
\end {eqnarray}
For large enough values of $\chi_l$.

The probability that a node of degree k will be outside layer No. $l+1$ is:

\begin{eqnarray}
P_{l+1}(k)= P_l(k)\*P(k,l+1|l)= 
\nonumber \\ = P_l(k)\*exp \left (-\frac{k}{1+\frac{T_l}{\chi_l}} \right)
\end {eqnarray}

Thus we derive the exponential cutoff:
\begin{equation}
\label {P_l(k)}
P_{l}(k)=P(k)\*exp \left(-\frac{k}{K_l} \right)
\end {equation}

Where :
\begin{equation}
\label {K_l}
\frac{1}{K_{l+1}}=\frac{1}{K_l}+\frac{1}{1+\frac{T_l}{\chi_l}}
\end {equation}

An alternate method for deriving the above relationship is given in Appendix A.

Now let us find the behavior of $\chi_l$ and $S_l$, where $S_l$ is the number of
links incoming to the $l+1$ layer (and approximately equals $N_{l+1}$, the
number of nodes in the $l+1$ layer). The number of
incoming connections to layer $l+1$ equals the number of connections
emerging from layer $l$, minus the number of connections looping
back into layer No. $l$. The probability for a connection to loop back
into layer $l$ is:
\begin{equation}
P(loop|l)=\frac{\chi_l}{\chi_l+T_l}
\end {equation}
and Therefore:

\begin{equation}
\label {S_l}
S_{l+1}=\chi_l \* \left(1-\frac{\chi_l}{\chi_l+T_l} \right)
\end {equation}

The number of connections emerging from all the nodes in layer
No. $l+1$ is $T_l - T_{l+1}$. This includes
also the number of incoming connections from layer $l$ into layer
$l+1$, which is equal to $S_{l+1}$. Therefore:

\begin{equation}
\label {chi_l}
\chi_{l+1}=T_l-T_{l+1}-S_{l+1}
\end {equation}

At this point we have the following relations: $T_{l+1}(K_{l+1})$
Eq. (\ref{T_l}) and Eq. (\ref{P_l(k)}), $S_{l+1}(\chi_l,T_l)$
Eq. (\ref{S_l}), $K_{l+1}(K_l,\chi_l,T_l)$ Eq. (\ref{K_l}), and
$\chi_{l+1}(T_l,T_{l+1},S_{l+1})$ Eq. (\ref{S_l}) and (\ref{chi_l}).
These relations may be solved numerically.  Note that approximate
analytical results for the limit $N \rightarrow \infty$ can be found
in \cite{CH02,DM02}.

\section{Empirical Results}

\begin{figure}
\epsfxsize=3in
\hskip 0.1in
\epsfbox{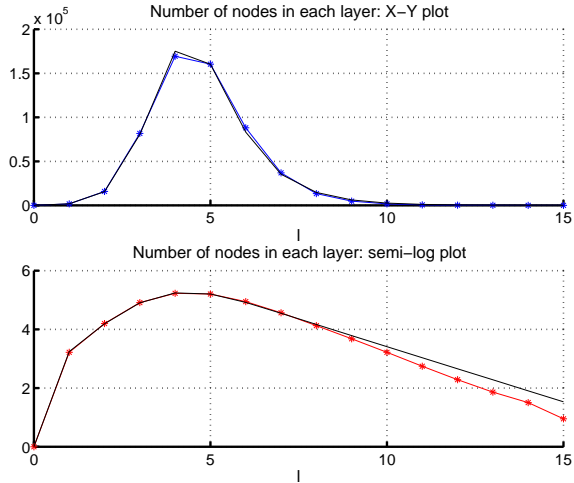}
\caption{\small $S_l$ vs. layer index $l$ for a network with $N=10^6$
nodes, $\lambda=2.85$, and $m=1$. Symbols represent simulation results
while black lines are a numerical solution for the derived recursive
relations. From the semi-log plot we see that there is an exponential
decay of $S_l$ for layers $l>L$ starting from a given layer L which we
believe is related to the radius of the graph.}
\label{fig_population}
\end{figure}

\begin{figure}
\epsfxsize=3in
\hskip 0.1in
\epsfbox{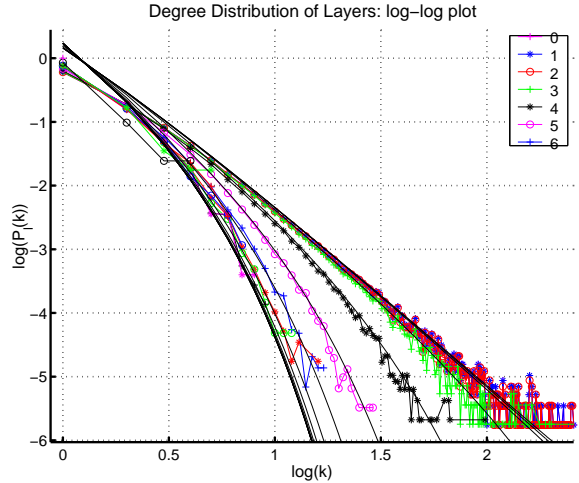}
\caption{\small log-log plot of $P_l(k)$ for different layers
  $l=0,1,2,...$, for a network with $N=10^6$ nodes, $\lambda=2.85$,
  and $m=1$. Symbols represent simulation results while black lines
  are a numerical solution for the derived recursive relations.}
\label{fig_distribution} \end{figure}

Figure \ref{fig_population} shows results from simulations (colored
symbols) for the number of nodes at layer $l$, which can be seen to be
in agreement with the analytical curves of $S_l$ (lines). We can see
that starting from a given layer $l=L$ the number of nodes decays
exponentially. We believe that the layer index $L$ is related to the
radius of the graph~\cite{CH02}. It can be seen that $S_l$ is a good
approximation for the number of nodes at layer $l$. This is true in
cases when only a small fraction of sites in each layer $l$ have more
than one incoming connection. An example for this case is when $m=1$
so that most of the sites in the network have only one
connection. Figure~\ref{fig_distribution} shows results for $P_l(k)$
with similar agreement.

It is important to note that the simulation results give the
probability distribution for the percolation cluster, while the
analytical reconstruction gives the probability distribution for the
whole graph. This explains the difference in the probability
distributions for lower degrees: many low degree nodes are not
connected to the percolation cluster and therfore the probability
distribution derived from the simulation is smaller for low degrees.

 \label{sec:net_anlz}

\begin{figure}
\epsfxsize=3in
\hskip 0.1in
\epsfbox{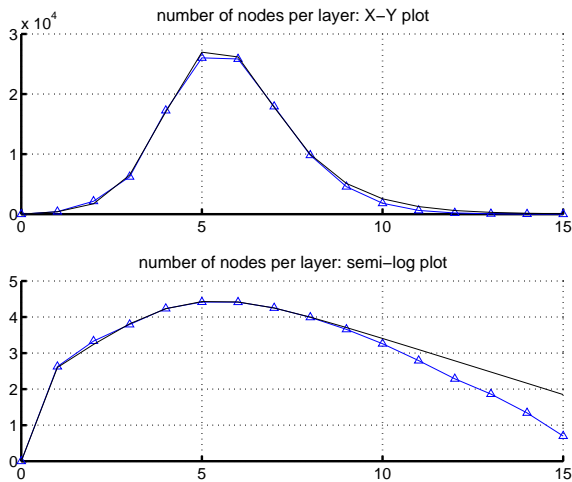}
\caption{\small number of nodes at each layer for a router level cut
of the internet with $N=112,969$ nodes (Lucent routers). Analytical reconstruction is
done with $\lambda=3$, and $m=1$.}
\label{fig_population_lucent}
\end{figure}

\begin{figure}
\epsfxsize=3in
\hskip 0.1in
\epsfbox{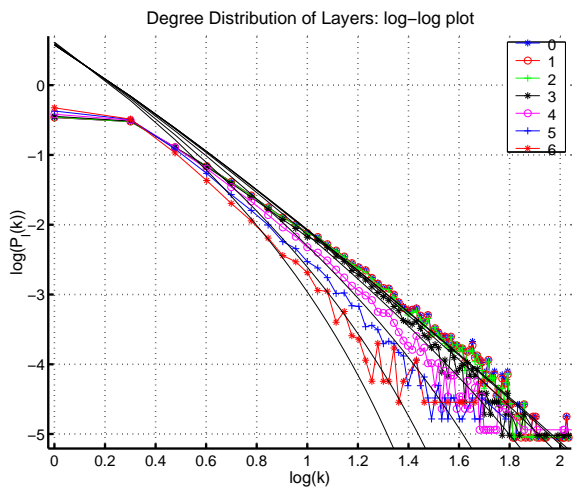}
\caption{\small log-log plot of $P_l(k)$ for different layers
  $l=0,1,2,...$, for a router level cut of the internet with
  $N=112,969$ nodes (Lucent routers). Qualitative analytical
  reconstruction is done with $\lambda=3$, and $m=1$.  }
\label{fig_distribution_lucent} 
\end{figure}

Figure \ref{fig_population_lucent} and figure
\ref{fig_distribution_lucent} show the same analysis for a cut of the
internet at router level (lucent routers). The actual probability
distribution is not a pure power law, rather it can be approximated by
$\lambda=2.3$ for small degrees and $\lambda=3$ at the tail. Our
analytical reconstruction of the layer statistics assumes $\lambda=3$,
because the tail of a power law distribution is the important factor
in determining properties of the system. This method results in a good
reconstruction for the number of nodes in each layer, and a
qualitative reconstruction of the probability distribution in each
layer.

In general, large degree nodes of the network mostly reside in the
lower layers, while the layers further away from the source node are
populated mostly by low degree nodes. This implies that the tail of
the distribution affects the lower layers, while the distribution
function for lower degrees affects the outer layers.  Thus the
deviations in the analytical reconstruction of the number of nodes per
layer for the higher layers may be attributed to the deviation in the
assumed distribution function for low degrees (that is: $\lambda=3$
instead of $\lambda=2.3$).

Our model does not take into account the correlations in node degrees,
which were observed in the internet ~\cite{N02}, and hierarchichal
structures ~\cite{VPV02}. This may also explain the deviation of our
measurments from the model predictions.

\label{sec:net_anlz}

\section{Empirical Findings on the Tomography of Multicast Trees}
\label{sec:trees}
In this section, we detail some of our findings on the structure and 
characteristics of the depth rings around the root node of shortest path trees. 
All of our findings were also validated on real Internet data.

\subsection{Topology and Tree Generation}
\label{sec:generation}
Our method for producing trees is the following. First, we generate power law
topologies based on the Notre-Dame model~\cite{AB00}. 
The model specifies 4 parameters: $a_0$, $a$, $p$ 
and $q$ \footnote{The notations in~\cite{AB00} are $m_0$, $m$, $p$ and $q$.}. 
Where $a_0$ is the initial number of detached nodes, and $a$ is the initial 
connectivity of a node.  When a link is added, 
one of its end points is chosen randomly, and the other with 
probability that is proportional to the nodes degree. 
This reflects the fact that new links
often attach to popular (high degree) nodes. 
The growth model is the following: with probability $p$, $a$ new links are 
added to the topology. With probability $q$, $a$ links
are rewired, and with probability $1-p-q$ a new node with $a$ links is added.
Note that $a$, $p$ and $q$ determine the average degree of the nodes.
We created a vast range of topologies, but concentrated on several parameter
combinations that can be roughly described as very sparse (VS),
Internet like sparse (IS) and less sparse (LS). 
Table~\ref{fig:topos} summarizes the
main characteristics of the topologies used in this paper.
 
\begin{table*}
\begin{center}
\begin{tabular}{|c|c|c|c|c|} \hline
Name & Type & Parameters & No. of Nodes & Avg. Node degree \\ \hline 
VS & generated & $a=1;p \in {0:0.05:0.5}$ & 10000 & $1.99 - 3.98$ \\ 
IS & generated & $a=2;p \in {0:0.05:0.5}$ & 10000 & $3.99 - 7.9$ \\ 
LS & generated & $a=3; p \in {0:0.05:0.5}$ & 10000 & $5.98 - 12.04$ \\ 
Big IS & generated & $a=1.5,2; p=0.1 $ & 50000;100000 & $3.3$,$4.4$ \\ 
BL[1,2] & real data & -- & Internet & 3.2 \protect\footnotemark \\ 
LC & real data & -- & Internet & 3.2 \protect\footnotemark \\ \hline
\end{tabular}
\caption{Type of underlying topologies used}
\label{fig:topos}
\end{center}
\end{table*}

From these underlying topologies, we create the trees in the following manner.
For each predetermined size of client population we choose 
a root node and a set of clients. Using Dijkstra's algorithm we build 
the shortest path tree from the root to the clients. 
To create a set of trees that realistically resemble Internet trees, we
defined four basic tree types. These types are based on the rank of the root
node and the clients nodes. The rank of a node is its location in a list of descending degree order, 
in which the lowest rank, one, corresponds to the node with the highest degree in the graph. 
For the case of a tree rooted at a big ISP site, we choose a root node with a low rank, thus ensuring
the root is a high degree node with respect to the underlying topology. Then, we either
choose the clients as high ranked nodes, or at random, as a control group. Note, that
due to the characteristic of the power law distribution,
a random selection of a rank has a high probability of choosing a low degree node. 
The next two tree types have a high ranked root, which corresponds to a multicast
session from an edge router. Again, the two types differ by the clients degree distribution,
which is either low, or picked at random.

The tree client population is chosen at the range $[50,4000]$ for the 10000 node generated topology,
$[50,10000]$ for the 100000 node generated topology, and 
$[500,50000]$ for the trees cut from real Internet data.
For each client population size, 14 instances were generated for each of
the four tree types.
All of our results are averaged over these instances. The variance of the results 
was always negligible. 
 
There are two underlying assumptions made in the tree construction.
The first, is that the multicast routing protocol delivers a packet from the source to each of 
the destinations along a shortest path tree.
This scenario conforms with current Internet routing. For example, IP packets are forwarded 
based on the reverse shortest path, and multicast
routing protocols such as Source Specific Multicast~\cite{SSM02} deliver packets along the 
shortest path route. 
In addition, we assume that client distribution in the tree is uniform, as has been shown 
by~\cite{PST99,CA01}.

\subsection{Tree Characteristics}
Our results show that trees cut from a power law topology obey a similar 
power law for the degree distribution, as well as the sub-trees 
sizes~\cite{DMS03}. The results were shown to hold for all trees cut from all generated topologies, even for trees as small as 200 nodes.

\begin{figure}[t]
\begin{center}
\pscript{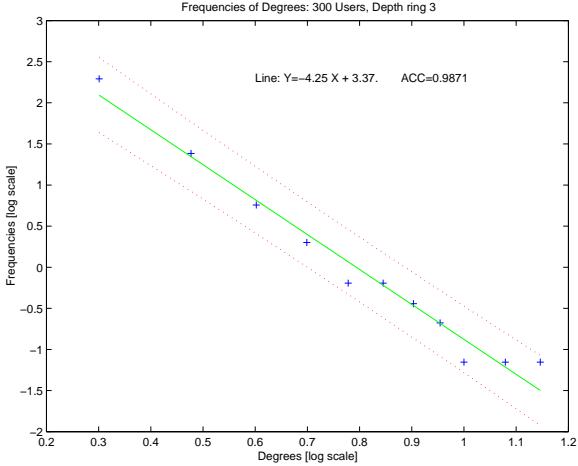}{0.95}
\caption{Third layer of a 300 client tree cut from topology ${a_0=6,a=1.5,p=0.1,q=0}$}
\label{fig:d_Bt6010_300R}
\end{center}
\end{figure}

\begin{figure}[t]
\begin{center}
\pscript{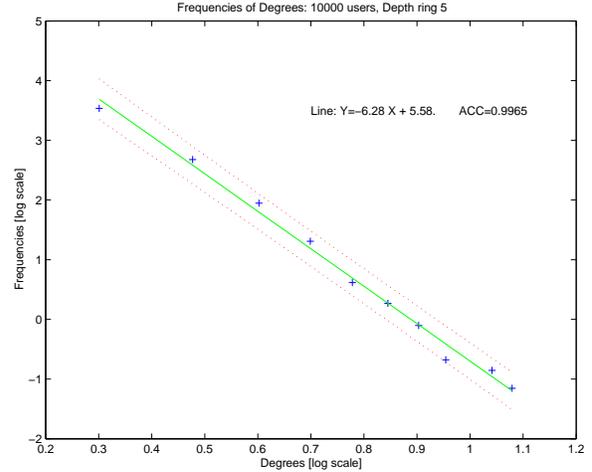}{0.95}
\caption{Fifth layer of a 10000 client tree cut from topology ${a_0=6,a=1.5,p=0.1,q=0}$}
\label{fig:d_Bt6010_10000R}
\end{center}
\end{figure}

In this work we further investigated the tomography of the trees, and looked at
the degree distribution of nodes at different depth rings around the root,
i.e., tree layers.
It was rather interesting to observe that any layer with sufficient number of nodes to create
a valid statistical sample obeyed a degree-frequency relationship which was similar to 
a power law, although with different slopes. We suspect that this is due to 
the exponential cut-off phenomenon discussed in the previous sections.
Figure~\ref{fig:d_Bt6010_300R} shows this for the third layer around the root (i.e.,
nodes at distance three from the root) of a 
300 client tree cut from a big IS topology (100000 nodes). 
The root was chosen with a high degree, and the clients with a low degree. 
Although the number of nodes is quite small, we see a very good fit with the power law.
Figure~\ref{fig:d_Bt6010_10000R} shows an excellent fit to the power law for the
fifth layer around the root of a 10000 client tree, cut from the same topology.
This phenomenon is stable regardless of the tree type, and the client population size.

\begin{figure}[t]
\begin{center}
\pscript{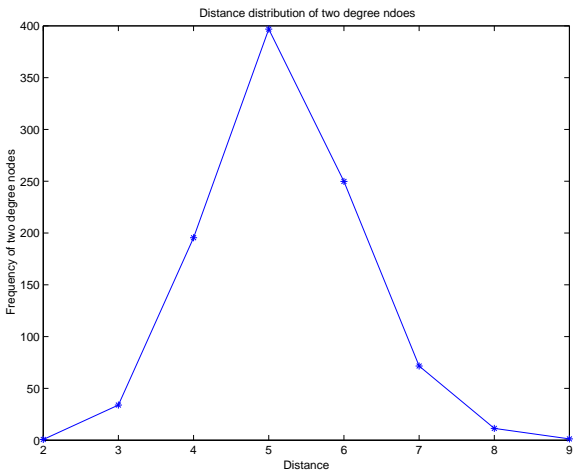}{0.95}
\caption{Distribution of degree two nodes in a tree cut from topology {$a_0=6,a=1,p=0.3,q=0$}} 
\label{fig:deg2_1000Rr}
\end{center}
\end{figure}

\begin{figure}[t]
\begin{center}
\pscript{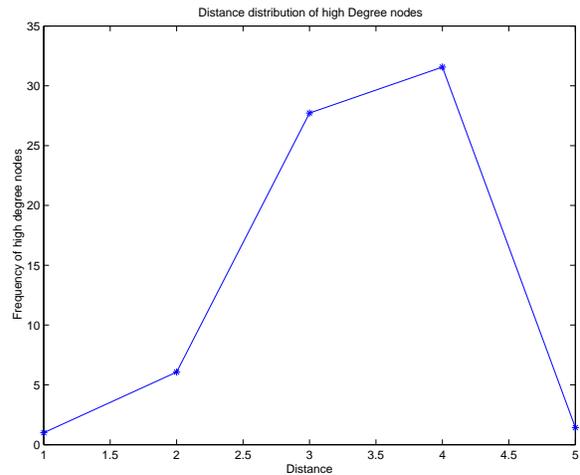}{0.95}
\caption{Distribution of the high degree nodes in a tree cut from topology {$a_0=6,a=1,p=0.3,q=0$}} 
\label{fig:degh_1000Rr}
\end{center}
\end{figure}

To understand the exact relationship of the degree-frequency at different layers, we
plotted the distribution of each degree at different layers.
Cheswick at al.~\cite{CNSSV99} found a gamma law for the number of 
nodes at a certain distance from a point in the Internet. Our results show
that the distribution of nodes of a certain degree at a certain distance (layer) from
the root seems close to a gamma distribution, although we did not determine its exact
nature.  Figure~\ref{fig:deg2_1000Rr} shows the distribution of the distance of 
two degree nodes, and Figure~\ref{fig:degh_1000Rr}  the
distribution of the distance of high degree nodes, i.e., nodes with a degree six and higher.
In both figures the root is a low degree node, and the tree has 1000 low degree clients.
As can be seen, the high degree nodes tend to reside much closer to the root than the low degree
nodes, and in adjacent layers.  In this example, most of them are in the second to forth 
layers around the root, with only two more at layer five.
This phenomenon was even more obvious when the root was a high degree node.

the distribution of the lengths of the paths to the clients. 
Our results show that the less connected the underlying topology, the 
higher is the average tree cut from the topology.
For a 10000 node underlying topology with an average degree of three and higher, 
the height of the trees was not more than ten. On an underlying topology of 100000 nodes,
the height of the trees was not more than 12. In accordance with our findings of a 'core'
of high degree nodes, the trees were higher on the average when the root was a low degree node,
compared to trees with a high degree root.


\begin{figure}[th]
\begin{center}
\pscript{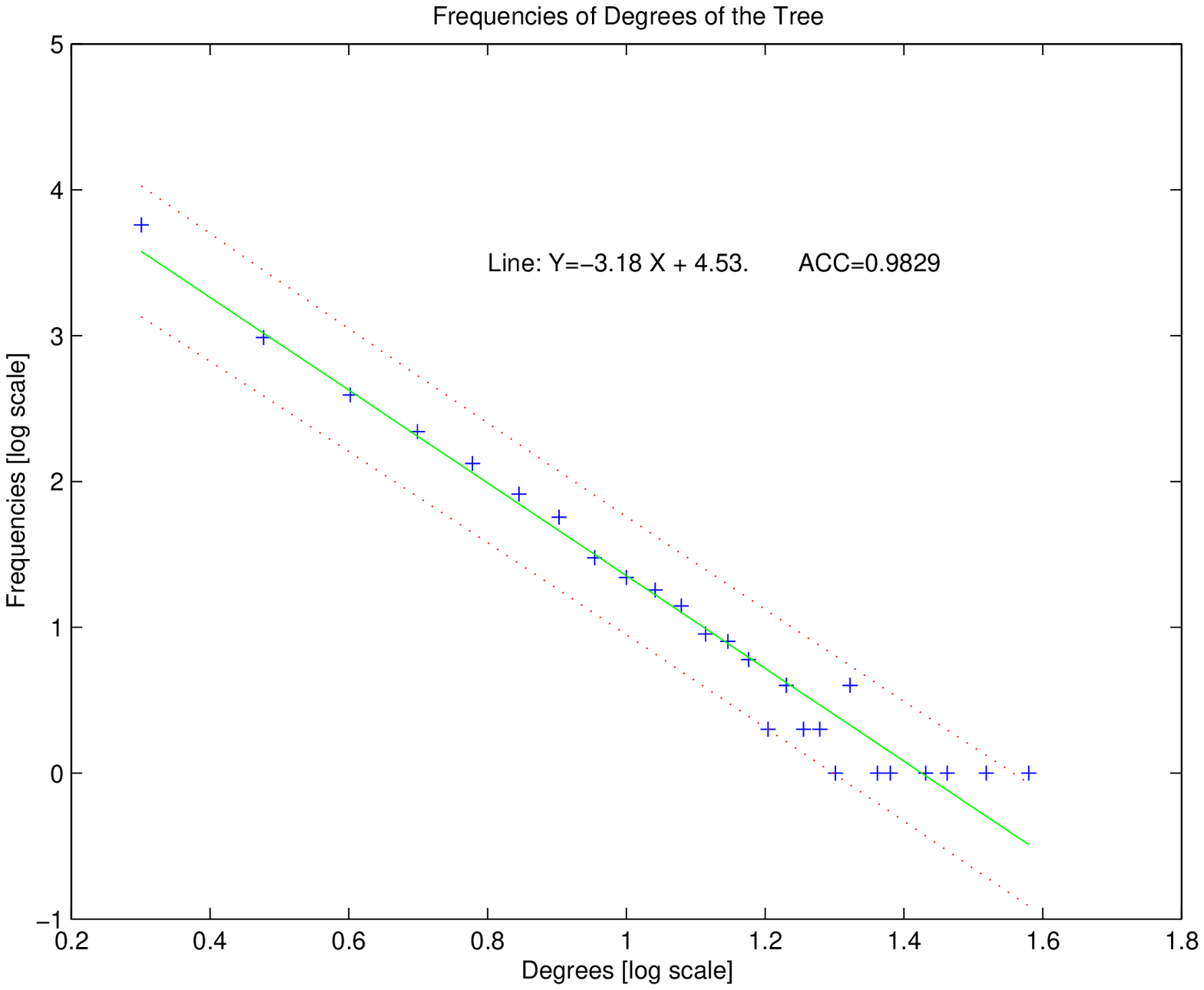}{0.95}
\caption{Frequency of degrees of the Internet tree.} 
\label{fig:blt0}
\end{center}
\end{figure}

\begin{figure}[th]
\begin{center}
\pscript{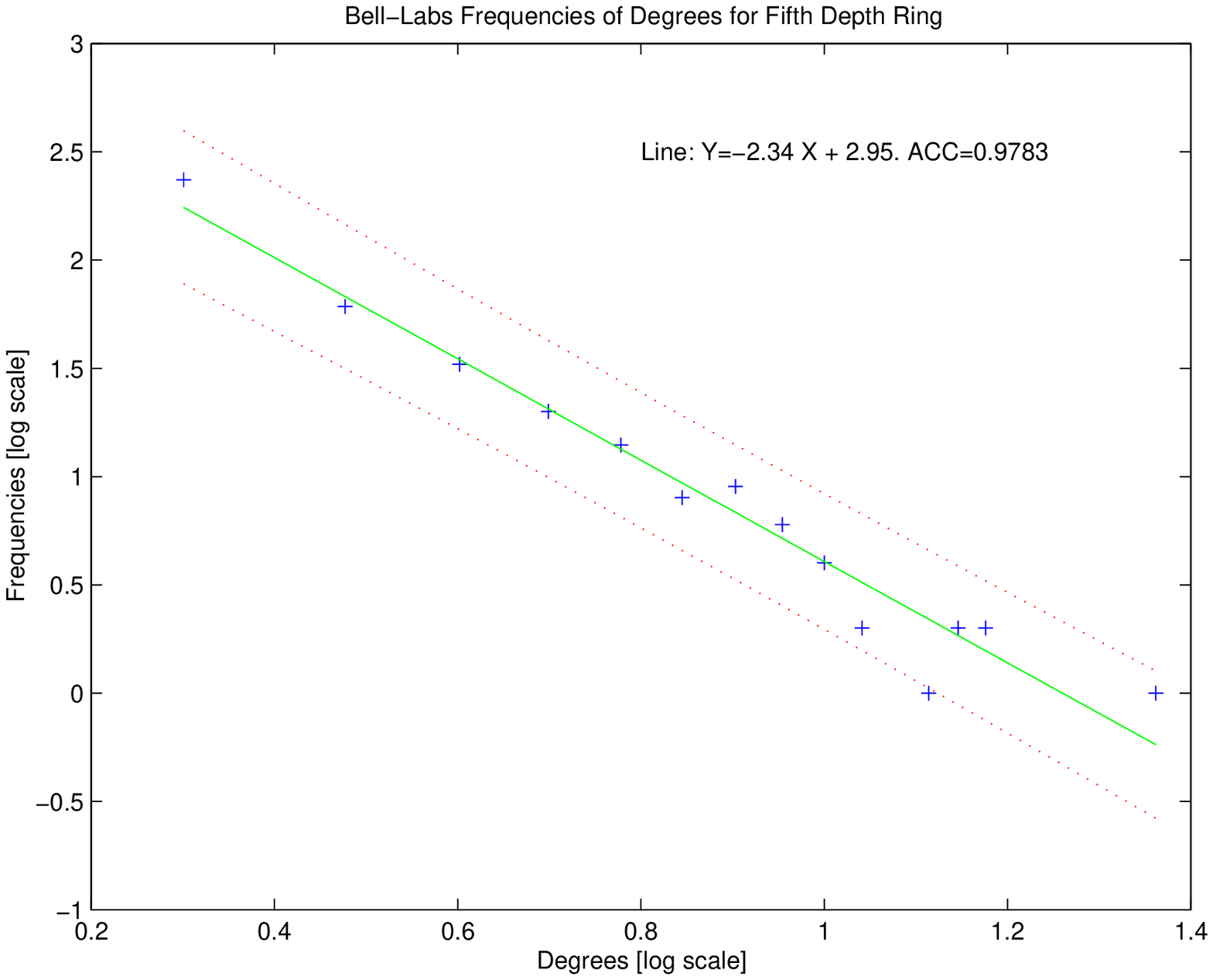}{0.95}
\caption{Frequency of degrees at layer 5 of the Internet tree.} 
\label{fig:blt0d5}
\end{center}
\end{figure}

\begin{figure}[ht]
\begin{center}
\pscript{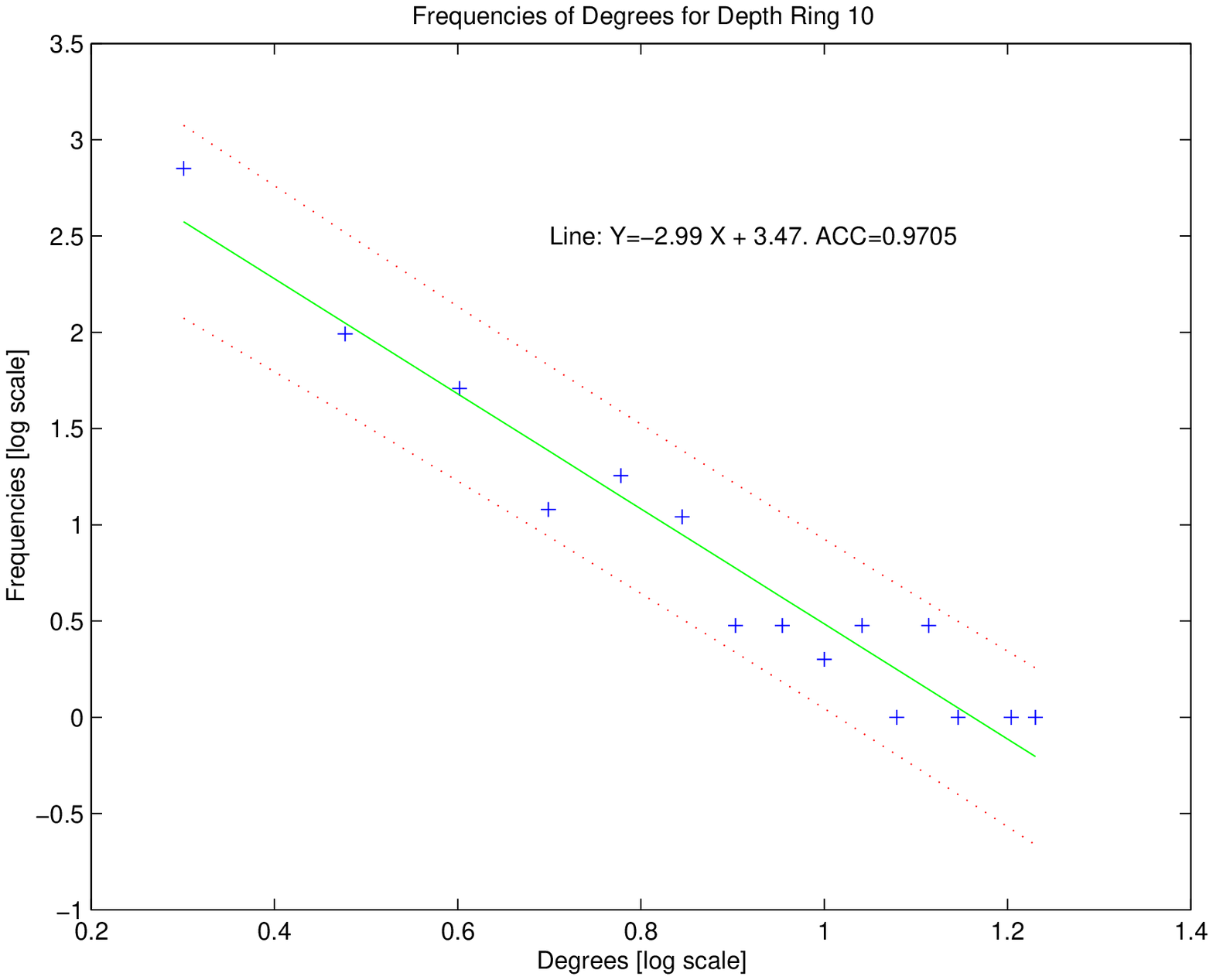}{0.95}
\caption{Frequency of degrees at layer 10 of the Internet tree.} 
\label{fig:blt0d10}
\end{center}
\end{figure}

We verify the above findings with results obtained from a real 
Internet data set.
Since we have no access to multicast tree data we use the client population of
a medium sized web site with scientific/engineering content.  This may represent
the potential audience of a multicast of a program with scientific content.
Two lists of clients were obtained, and traceroute was
used to determine the paths from the root to the clients. 
It is important to note, that the first three levels of the tree
consist of routers that belong to the site itself, and therefore might be treated as the root 
point of the tree, although in these graphs they appear separately.
Figure~\ref{fig:blt0} shows the frequency of degrees in the tree. The linear fit of the log-log
ratio is excellent, with a correlation coefficient of 0.9829. The exponent is very close to
the exponent we derived for trees cut from topologies that resemble the 
Internet.
Figures~\ref{fig:blt0d5}~and~\ref{fig:blt0d10} show the frequency of degrees
at layers 5 and 10 of the tree, respectively. They conform with our finding that
the power law of frequency-degree is maintained for each separate distance around the root.

\section{Conclusions}
\label{sec:conc}

We define a ``layer'' in a network as the set of nodes at a given
shortest path distance from a chosen node. We find that the degree
distribution of the nodes at each layer obeys a power law with an
exponential cutoff. We also model the behavior of the number of nodes
at each layer, and explain the observed exponential decay in the outer
layers of the network.


\label{sec:net_anlz}


\renewcommand{\baselinestretch}{0.9} 
\small
\bibliographystyle{abbrv}
\bibliography{layers}  
\clearpage
\appendix
\noindent
\begin{center}
\textbf{\LARGE Appendix A} \\
\label{app:appx}
\textbf{Deriving $P_l(k)$ Using Mean Field Approximation}
\end{center}

Each node is treated independently, where the
{\em interaction} between nodes is inserted through the expected number
of incoming connections. At each node, the process is treated as
equivalent to randomly distributing $\chi_l$ independent points on a
line of length $\chi_l+T_l$ and counting the resultant number of
points inside a {\em small} interval of length $k$. Thus, the number
of incoming connections $k_{in}$ from layer $l$ to a node with $k$
open connections is distributed according to a Poisson distribution
with:
\begin{equation}
\label {k_in_average}
<k_{in}>=\frac{k}{\chi_l+T_l}\chi_l 
\end {equation}

and :

\begin{equation}
\label {k_in_distribution}
P_{l+1}(k_{in}|k)=e^{-<k_{in}>} \frac{<k_{in}>^{k_{in}}}{k_{in}!}
\end {equation}

The probability for a node with $k$ open connections {\em not} to be
connected to layer $l$, i.e. to be outside layer $l+1$ also, is:

\begin{eqnarray}
\label {k_in_conditional_prob}
P(k,l+1|l)=P_{l+1}(k_{in}=0|k)=e^{-<k_{in}>}=
\nonumber \\ =exp \left(-\frac{k}{1+\frac{T_l}{\chi_l}} \right)
\end {eqnarray}

Thus the total probability to find a node of degree $k$ outside layer $l+1$ is: 
\begin{equation}
\label {k_in__prob}
P_{l+1}(k)=P_l(k)P(k,l+1|l)=P_l(k)exp \left(-\frac{k}{1+\frac{T_l}{\chi_l}} \right)
\end {equation}

And we receive an exponential cutoff.

\end{document}